\documentclass[aps,prb,twocolumn,showpacs,superscriptaddress,10pt]{revtex4-1}\begin{tiny}\end{tiny}
\usepackage{longtable}
\usepackage{amsfonts}
\usepackage[pdftex]{graphicx}
\usepackage{rotating}
\usepackage{hyperref}
\usepackage{dcolumn}
\newcolumntype{d}{D{.}{.}{3.5}}

\begin{document}

\title{Competition between exchange-driven dimerization and magnetism in diamond(111)}

\author{Bet\"{u}l Pamuk}
\email{betul.pamuk@cornell.edu}
 \affiliation{School of Applied and Engineering Physics, Cornell University, Ithaca, NY 14853, USA}
 
\author{Matteo Calandra}
 \email{matteo.calandra@upmc.fr}
\affiliation{Sorbonne Universit\'e, CNRS, Institut des
  Nanosciences de Paris, UMR7588, F-75252, Paris, France}

\date{\today}


\begin{abstract}
Strong electron-electron interaction in ultraflat edge 
states can be responsible for correlated phases of matter, such as
magnetism, charge density wave or superconductivity. 
Here we consider the diamond(111) surface that,
after Pandey reconstruction, presents zig-zag carbon
chains, generating a flat surface band. 
By performing full structural optimization with hybrid functionals and neglecting spin polarization, 
we find that a substantial dimerization 
($0.090$\,\AA / $0.076$\,\AA\,  bond disproportionation in the PBE0/HSE06) 
occurs on the chains; 
a structural effect absent in calculations based on the LDA/GGA functionals.
This dimerization is the primary mechanism for the opening of an
insulating gap in the absence of spin polarization. 
The single-particle direct gap is $1.7$ eV ($1.0$ eV) in the PBE0 (HSE06), 
comparable with the experimental optical gap of $1.47$ eV, 
and on the larger(smaller) side of the estimated experimental 
single particle gap window of 1.57-1.87 eV, after inclusion of excitonic effects. 
However, by including spin polarization in the calculation, we find
that the exchange interaction stabilizes a different ground state,
undimerized, with no net magnetization and ferrimagnetic along the
Pandey $\pi$-chains with magnetic moments as large as $0.2-0.3~\mu_B$ in the PBE0.
The direct single-particle band gap in the equal spin-channel 
is approximately $2.2$ eV ($1.5$ eV) with the PBE0 (HSE06) functional.
Our work is relevant for systems with flat bands in general and wherever the
interplay between structural, electronic and magnetic degrees of freedom is
crucial, as in twisted bilayer graphene, IVB atoms on
IVB(111) surfaces such as Pb/Si(111) or molecular crystals.
\end{abstract}


\maketitle


\section{Introduction}

The occurrence of strongly correlated states requires the dominance of
the electron-electron interaction over the electronic kinetic
energy. In the case of 3$d$ transition metal oxides or high $T_c$
superconductors, Mott insulating, magnetic, and superconducting states
are stabilized via the strong localization of electrons in 3$d$ orbitals.
Recently, it has been shown that a new class of strongly correlated
systems can be achieved in ultraflat edge-states or surface bands
having small Fermi velocities but not necessarily 3$d$ states,
as it happens in twisted bilayer graphene\cite{CaoMott,CaoSC},
or multilayer graphene with rhombohedral stacking 
\cite{Lau2016,Thomas2015,Faugeras2016,Henck_Rhombo,BetulRhombo2017}.
All these works expand the range of materials hosting strong correlation
effects and exotic states of matter and point to the need of
understanding exchange and correlation effects in flat edge-states.

\begin{figure}[!thb]
        \centering 
                \includegraphics[clip=true, trim=0mm 0mm 0mm 0mm, width=0.48\textwidth]{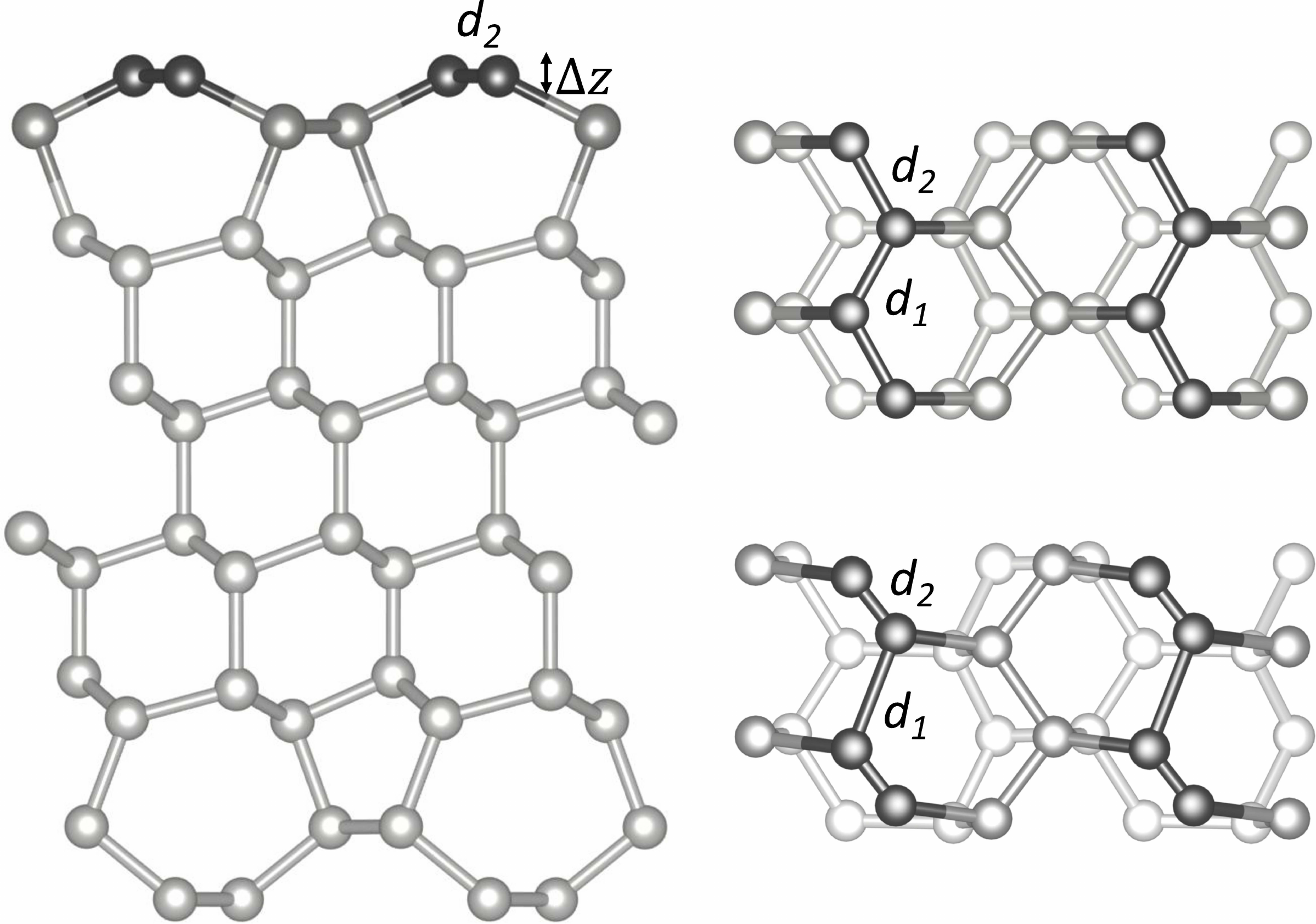}
        \caption{The structure of the diamond surface along the [111]
          direction plotted in a $2\times2$ cell. 
        Left: Side view with 12 layers of C atoms. The buckling of the top layer is $\Delta z$. 
        Right: Above is the undimerized, and below is the dimerized top view with top 4 layers shown.  
        The black atoms are the topmost layer, dark gray atoms are the second layer, and light gray atoms are the rest of the layers. 
        The bond lengths of the topmost layers are labeled $d_1$ and $d_2$,
        showing the change in the bond length of the top layer with inclusion of the exact exchange. 
        }
        \label{fig:struc}
\end{figure}

One of the simplest and most studied systems hosting a flat edge-state prone to
strong exchange-correlation effects is the diamond(111) surface -- the structure of which
is still under debate more than 100 years after Bragg got the Nobel prize and applied their
diffraction technique to determine the structure of bulk diamond \cite{Bragg1913}. 
The formation of the surface state can be understood by considering that in bulk diamond
each carbon atom undergoes $sp^3$ hybridization and
has four neighbors at distance of $\approx 1.54$\,\AA\, and bond
angle at 109.5$^o$. The atoms on the (111) surface have,
however, one missing bond and only three nearest neighbors. 
This dangling bond generates the so-called
Pandey \cite{Pandey1982} reconstruction resulting in a $2\times 1$
superstructure forming 1D zig-zag chains (see Fig. \ref{fig:struc}) and a surface electronic
band. The weak, but not negligible, hopping integral in the direction parallel to the surface but perpendicular to the
chains is responsible for the $\approx 0.5$ eV energy dispersion of
the band.
Even if this surface state is not as flat as the one detected in
twisted bilayer graphene\cite{CaoMott,CaoSC} or in multilayer graphene
with
rhombohedral stacking
\cite{Lau2016,Thomas2015,Faugeras2016,Henck_Rhombo,BetulRhombo2017},
it is substantially more extended in reciprocal space and it holds a
larger number of electrons. For this reason diamond(111) should be
prone
to strong exchange and correlation effects.

Despite its apparent simplicity, the theoretical and experimental description
of the structural and ground state properties of the diamond(111)
surface has proven to be an exceptionally difficult and yet
unsolved problem. 

Although it is well accepted (both in theory and experiments) 
that the surface is a Pandey $\pi$-chain \cite{Pandey1982},
the microscopic details, such as dimerization, buckling, deeper layer
distortions are still under debate.
While X-ray diffraction \cite{vanderVeen1998} 
and ion scattering \cite{vanderVeen1998b} 
data suggest buckling of the surface atoms,
low-energy electron diffraction measurements \cite{Ley2002} show that the buckling
is negligible, but the dimerization is inconclusive within the experimental error.
For the electronic structure,
angle-resolved photoemission spectroscopy (ARPES) \cite{Stampfl1997ARPES} measurements find
an insulating state with the occurrence of the flat surface state 0.5 eV below the Fermi level.
Electron energy loss spectroscopy \cite{Pepper1982EELS} measurements suggest a band gap of $\sim 1$ eV, and
reflectance anisotropy spectroscopy \cite{Derry2007RAS,Derry2009RAS} 
gives larger values of the optical band gap of $\approx 1.47$ eV. 

From the theoretical point of view, the first density functional theory (DFT) calculations 
within the standard local density (LDA) and generalized gradient (GGA) approximations
gave conflicting results for the surface chains \cite{Vanderbilt1984, Galli1992, Alfonso1995}.
This disagreement is most likely explained by the fact that calculations
were very heavy for the time as 12 carbon layers are required for convergence.
More recent calculations with the LDA and GGA functionals 
\cite{Kresse1996,Bechstedt1996,Bechstedt2001,Stekolnikov2002,Marsili2005,Marsili2007},
do not present any buckling or dimerization of the structure.
Similarly a calculation with the B3LYP functional also does not show buckling or dimerization \cite{DeLaPierre2014}.
Consequently, the surface state appears to be metallic within
the standard LDA/GGA approximations, in disagreement with all experimental
data\cite{Stampfl1997ARPES, Bechstedt1996,Pepper1982EELS,Derry2007RAS,Derry2009RAS}.
Pioneering works by Marsili {\it et al.}
\cite{Marsili2005,Marsili2007,Marsili2008}
show that quasiparticle GW calculations on top of the GGA crystal structure
lead to a gap opening only within the self-consistent $G_1W_1$ scheme
and starting from an artificial band occupation. 
However, GW calculations were performed with very
coarse grids (5 or 9 k-points in the Brillouin zone) that tend to substantially
overestimate the band gap (see Appendix \ref{app:kpnt}), and in the absence
of spin polarization. In these works, the
band gap opening is attributed only to an electronic mechanism at fixed ionic
coordinates. Finally, the magnitude of excitonic effects has been
evaluated to be of the order of $0.1-0.4$ eV \cite{Marsili2008}
leading to an experimental single particle gap of the order of $1.57-1.87$ eV,
when added to the experimental gap of 1.47 eV.

The main problem of all previous theoretical works is that they rely on
the GGA minimized structure, mostly because structural optimization
within GW is not possible for solids and non-local exchange calculations in
a plane-wave framework are too expensive for such a large system (24 atoms per
cell and very dense electronic momentum k-point mesh). Moreover, all 
calculations neglected magnetism.
In this work, we circumvent the difficulties of geometrical optimization with a
dense mesh of electronic momentum k-points even in the existence of
Hartree-Fock exchange by using a combination of plane waves\cite{QE,QE-2017}  and Gaussian basis
sets \cite{Crystal14}  that allow for fast structural optimization.
We use hybrid functionals with exact exchange and
range separation to understand the effects of
the exchange interaction on the geometry 
and electronic structure of diamond(111).
Finally, we also explore the occurrence of magnetic solutions.

The structure of the paper is the following: after explaining
technical
details in sec. \ref{sec:tec}, we present results for
non-magnetic (sec. \ref{sec:nonmag}) and magnetic (sec. \ref{sec:mag})
calculations. 

\section{Technical details \label{sec:tec}}

DFT calculations are performed using the {\sc Quantum ESPRESSO} \cite{QE,QE-2017} and CRYSTAL
codes\cite{Crystal14}.
We use the
triple-$\zeta$-polarized Gaussian type basis sets for the C atoms \cite{Cbasis},
with the PBE \cite{PBE}, PBE0 \cite{PBE0}, and HSE06 \cite{HSE06} functionals.
The surface states require an ultradense sampling
with an electronic momentum k-point mesh of $90\times120\times1$, crucial
for an accurate determination of the band gap (as shown in Appendix \ref{app:kpnt}) and for the stabilization of magnetism.
We used real space integration tolerances of 7-7-7-15-30,
and an energy tolerance of $10^{-10}$ Ha for the total energy convergence.
Fermi-Dirac smearing for the occupation of the electronic states of 0.001 Ha
is used for all of the calculations.
In the magnetic case, we further increase the energy tolerance to $10^{-11}$ Ha.
We fix the magnetic state in the first iteration
of the self-consistent cycle and then we release this constraint.

\section{Results}

\subsection{Non-magnetic calculations \label{sec:nonmag}}

Fig. \ref{fig:struc} shows the diamond structure along the [111] direction in a $2\times2\times1$ cell. 
We consider 12 carbon layers with the bottom layer saturated by
hydrogen and the top unhydrogenated. 
We choose an in-plane lattice parameter of $a=4.369$ \AA, and $b=2.522$ \AA,
as derived from the experimental lattice constant of bulk diamond
$a_0=3.567$ \AA~ \cite{vanderVeen1998},
but we also perform full structural optimization (cell and internal
coordinates) although the results are weakly affected.
A vacuum of 50 \AA\, is placed between the periodic images along the
$z$-direction.
We first optimize the structure within the PBE using both plane waves and
Gaussian basis sets, finding practically indistinguishable results.
We then use the PBE optimized structure as a starting guess for
geometrical optimization with hybrid functionals. In the Gaussian
basis set calculations we do not use any symmetry so that no {\it a
  priori} guess on the crystal structure is retained. 

The Pandey $\pi$-chain surface atomic structure can be parametrized by
 the dimerization $\Delta$ and the
buckling $\Delta z$
(see Fig. \ref{fig:struc}),where 
$d_1$ and $d_2$ label the two distinct bond lengths of the atoms that make up the zigzag chain on the topmost layer.
The dimerization is then defined as: 
$\Delta=|d_1-d_2|/(d_1+d_2)$.
The buckling of the atoms, $\Delta z$, on this layer is simply the difference in their position
along the $z$-direction (see Fig. \ref{fig:struc}).

\begin{table}[!htb] \footnotesize
\caption{Bond lengths ($d_1$ and $d_2$) of the atoms in the Pandey $\pi$-chains in the
  topmost layer, dimerization $\Delta$, and buckling
  $\Delta z$ for different exchange and correlation functionals (XC).}
\centering
\begin{ruledtabular}
        \begin{tabular}{l c c c c}
        XC    & $d_1$(\AA)& $d_2$(\AA) & $\Delta$ & $\Delta z$ (\AA) \\
        \hline
        PBE   & 1.440 & 1.440 & 0.000 & 0.0048 \\
        HSE06 & 1.476 & 1.400 & 0.026 & 0.0052 \\
        PBE0  & 1.483 & 1.393 & 0.031 & 0.0054 \\
        B3LYP & 1.482 & 1.396 & 0.030 & 0.0048 \\
        \end{tabular}
        \end{ruledtabular}
\label{table:struc}
\end{table}

Table \ref{table:struc} shows the calculated values for the topmost
layer for the optimized atomic structures.
With all three functionals, the buckling of the top layer is estimated to be very small,
with $\Delta z \sim 0.005$ \AA,
which agrees well with the low-energy electron diffraction measurements of about 0.01 \AA~ \cite{Ley2002}.
With the PBE functional, the two bond lengths are equal, 
$d_1=d_2$, hence there is no dimerization.
The inclusion of unscreened exchange with the PBE0 functional,
or of screened exchange via the HSE06 functional, 
gives a significant imbalance between the two bond lengths, 
predicting a dimerization of the surface structure with $\Delta \sim
0.026$ in the HSE06 case.
The dimerization is slightly larger with the PBE0 functional than with
the HSE06 functional. In general, we find that 
the larger the amount of Hartree-Fock exchange included in the
calculation and the more unscreended the exchange, the larger the dimerization.
The energy gain induced by the dimerization is substantial, as shown
in Table \ref{table:deltaE}.

\begin{table}[!htb] \footnotesize 
\caption{The energy difference between the dimerized and undimerized
  structures and between magnetic and non-magnetic structures using
  different hybrid functionals. We use for the undimerized structure
 the PBE one, as the HSE06 and PBE0 functionals do not have a stable
 undimerized solution. We then obtain its energy in the HSE06 and PBE0
at fixed atomic positions.
The magnetic structure is fully optimized and has no dimerization. The non-magnetic
is also completely optimized and has dimerization. }
\centering
\begin{ruledtabular}
        \begin{tabular}{l c c c}
		$\Delta$ E (eV/cell) &  HSE06  & PBE0 & B3LYP \\
\hline
dimerized $-$ undimerized & -0.042      &  -0.062 & -0.022 \\
magnetic $-$ non-magnetic  &  -0.007     &  -0.008  & -0.002 \\
	    \end{tabular}
 \end{ruledtabular}
\label{table:deltaE}
\end{table}

In a previous calculation using similar settings and the B3LYP
functional, the author of Ref.  \onlinecite{DeLaPierre2014} found no
dimerization in the carbon chains. We repeated this calculation
starting (i) from the undimerized PBE structure and (ii) from the dimerized
HSE06 structure. In the first case, the simulation remains in the
undimerized structure as in Ref.  \onlinecite{DeLaPierre2014}. However,
in the second case, the structural optimization with the B3LYP
functional converges to a dimerized structure that is lower in energy
than the undimerized one, on the same line of what has been obtained with the 
other hybrid functionals.

\begin{figure*}[!t]
        \centering 
                \includegraphics[clip=true, trim=75mm 20mm 85mm 20mm, width=1\textwidth]{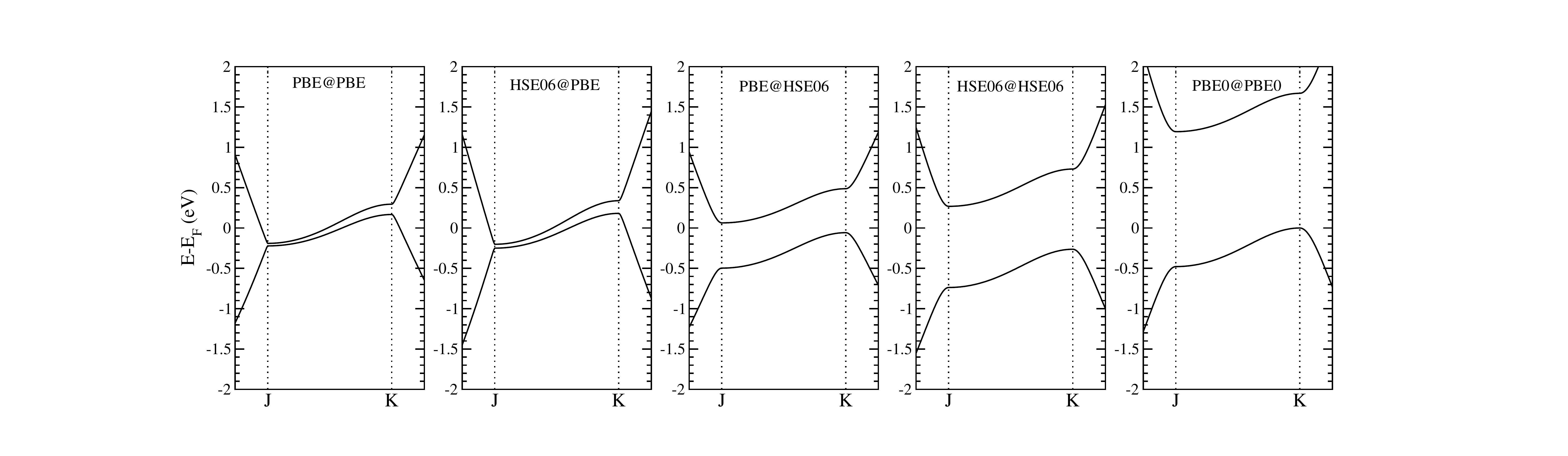}
        \caption{The electronic structure of the diamond(111) surface
          with different approximations. The notation
          functional1@functional2 means that the calculation of the
          electronic structure is
          performed using functional1 but with the crystal structure
          obtained by geometrical optimization using the functional2.}
        \label{fig:allBands}
\end{figure*}

Having demonstrated the crucial effect of the exchange interaction on
the atomic structure, we now investigate its effect on the electronic
spectrum. We first consider the PBE approximation on top of the PBE 
optimized structure (labeled PBE@PBE). We find, in agreement with all
previous calculations
\cite{Kresse1996,Bechstedt1996,Bechstedt2001,Stekolnikov2002,Marsili2005,Marsili2007}, 
a metallic solution with no gap and practically
very small direct gaps, as shown in Fig. \ref{fig:allBands}. Interestingly, the use of the HSE06
 on top of the PBE crystal structure (HSE06@PBE) still leads to a
 metallic solution with no indirect gap and tiny direct gaps at J and K.  On the contrary, if
 the PBE functional is used on top of the HSE06 geometry (PBE@HSE06), a gap opens
 and the electronic structure is insulating with an indirect gap of
 0.121 eV and direct gaps at J and
 K of 0.560 eV and 0.545 eV, respectively.
The fact that HSE06 on top of the PBE structure
leads to a metallic insulating solution, while even the
PBE functional on top of the dimerized HSE06 geometry is successful in
inducing an insulating state, demonstrates unambiguously that, in the
absence of spin polarization, gap opening is
mostly driven by the dimerization of the 
Pandey $\pi$-chains. Contrary to all previous works that
tried to stabilize an insulating solution at fixed atomic
coordinates\cite{Marsili2005,Marsili2007,Marsili2008}, our work underlines
the crucial importance of the atomic distortion.
The complete HSE06 calculation (i.e. HSE06 on top of the HSE06 structure,
HSE06@HSE06) leads to a larger direct gap of $\approx 1$ eV (see
Table \ref{table:Eg}) both at K and J  and  to a fundamental indirect gap
of $\approx 0.532$ eV. 
The electronic bands of the full Brillouin zone
as well as a comparison with the ARPES data are given in Appendix \ref{app:arpes}.
A larger direct gap of $1.7$ eV can be obtained using
unscreened functionals such as PBE0, as shown Fig. \ref{fig:allBands}.
The direct band gaps at different high-symmetry points and for all the 
used approximations are reported in Table \ref{table:Eg}.

\begin{table}[!htb] \footnotesize 
\caption{For each electronic band structure (obtained with XC) calculated @ the atomic structure (relaxed with XC), 
			the fundamental band gap $E_g$, and the direct band gap at the high-symmetry points of the 
			Brillouin zone, J, K, $\Gamma$, J', given in eV. 
                }
\centering 
\begin{ruledtabular}
        \begin{tabular}{l l c c c c c }
        bands & @ structure           & $E_g$ & J     & K     & $\Gamma$ & J' \\
        \hline 
	PBE   & @ PBE                 & 0.000 & 0.031 & 0.128 & 4.392 & 5.279 \\
	HSE06 & @ PBE                 & 0.000 & 0.048 & 0.157 & 5.562 & 6.607 \\
	PBE   & @ HSE06               & 0.121 & 0.560 & 0.545 & 4.438 & 5.322 \\
	HSE06 & @ HSE06               & 0.532 & 1.006 & 0.994 & 5.606 & 6.650 \\
	PBE0  & @ PBE0                & 1.194 & 1.672 & 1.670 & 6.339 & 7.381 \\
	B3LYP & @ B3LYP               & 0.961 & 1.421 & 1.407 & 5.925 & 7.032 \\
        \end{tabular}
        \end{ruledtabular}
\label{table:Eg}
\end{table}

As hybrid functionals give only a slight underestimation of the
experimental band gap in bulk
diamond\cite{Kummer2009,Garza2016}, the $32\%$ ($0.47$ eV)
underestimation in HSE06 of 
the optical gap with respect to the experiments
\cite{Derry2007RAS,Derry2009RAS}
is fairly surprising. The situation is much better in the PBE0 leading to a
somewhat larger gap than the optical direct gap measured in 
experiments  ($0.1$ eV larger).
However, this value is in better agreement with experiments;
if we add the excitonic effects of $0.1-0.4$ eV \cite{Marsili2008} 
to the experimental gap of 1.47 eV, 
this leads to an experimental single particle gap of $1.57-1.87$ eV. 
The PBE0 value is thus on the lower side of the window ($1.7$ eV). 
However, all calculations presented up to now have been carried out neglecting
spin-polarization. The occurrence of a flat band could also
lead to magnetic solutions\cite{CaoMott,CaoSC}, as it happens in
multilayer graphene with rhombohedral
stacking\cite{Henck_Rhombo,BetulRhombo2017},
a very similar system. For this reason, we
investigate below the occurrence of magnetism using the HSE06 and PBE0 functionals.

\subsection{Magnetic calculations  \label{sec:mag}}

\begin{figure*}[!thb]
        \centering
                \includegraphics[clip=true, trim=0mm 0mm 0mm 0mm, width=0.75\textwidth]{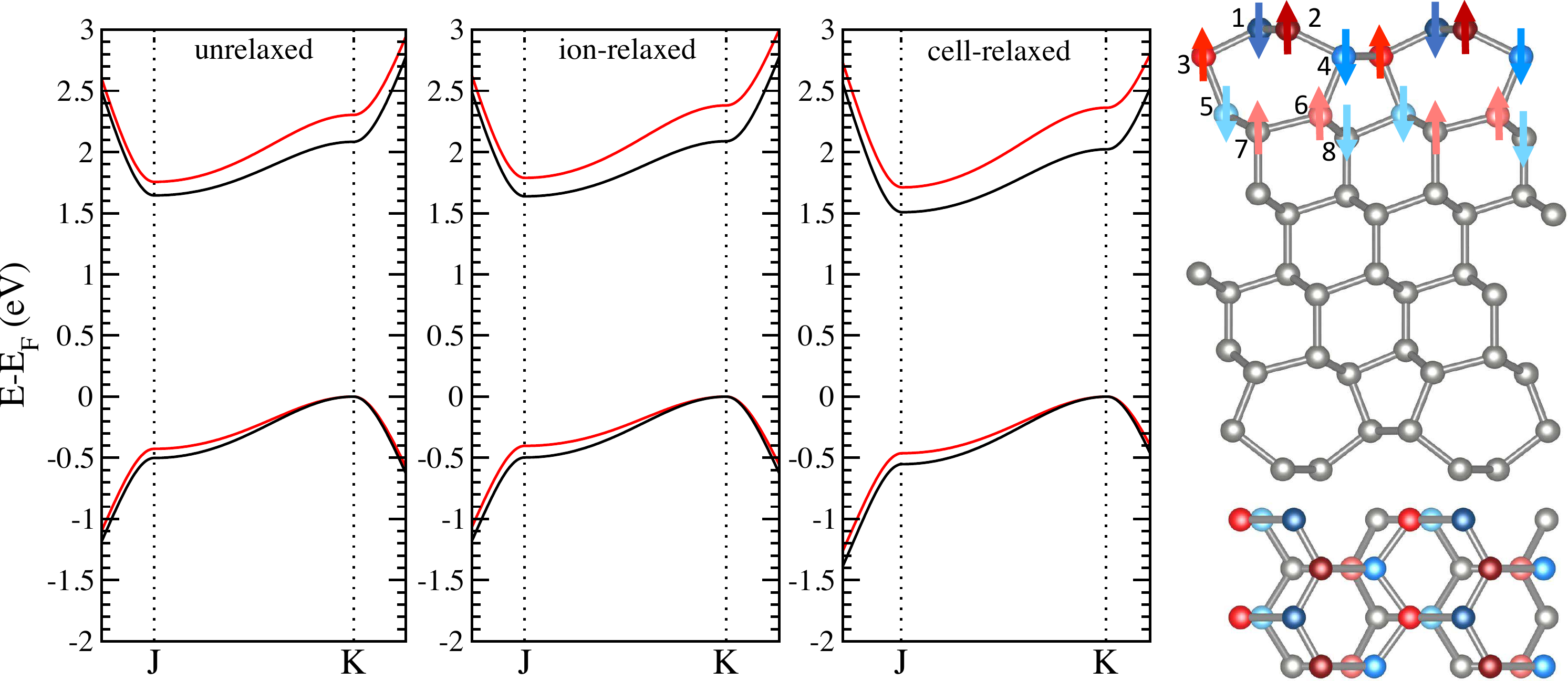}
        \caption{Left: The effect of magnetism on the electronic structure of diamond C(111)2$\times$1 surface 
        calculated with the PBE0 functional, with the unrelaxed (dimerized) structure of the non-magnetic calculation,
        and with the ion-relaxed and cell-relaxed calculations.
        Red is $\alpha$ (majority) spin and black is the $\beta$ (minority) spin electrons.
        Valence band maximum of each band is set to 0 eV.
        Right: The spin on each atom.
        Blue atoms are spin up and red atoms are spin down. 
        The decrease in the shade of the color denotes the decrease in the magnitude of the atomic spin going towards the bulk.        
        The arrows show the direction only and are not to scale. 
        The magnitude of each spin is given in Table \ref{table:spin}.
        }
        \label{fig:magBands}
\end{figure*}

We first perform magnetic calculations at atomic coordinates fixed at
the dimerized solution, referred to as ``unrelaxed" in the remainder of the figures and tables.
We choose as initial condition of the
simulation a fully ferromagnetic (FM) and antiferromagnetic (AFM) configuration
on the two surface atoms of the topmost layer
along the zig-zag chain. 
As expected, the PBE functional does not stabilize a magnetic state, therefore
we focus on the PBE0 and HSE06 functionals.
In both cases, we find that the most stable solution
has global zero magnetic moment, and is ferrimagnetic within each layer of atoms with large atomic magnetic moments.
With the HSE06, the magnetic moments on the two atoms
in the chain are $+0.271~\mu_B$ and $-0.269~\mu_B$. 
A similar solution is obtained with the PBE0, namely we obtain 
an atomic spin of $+0.288~\mu_B$ and $-0.285~\mu_B$ left on the atoms.
The small imbalance of $0.003~\mu_B$ between the
majority and minority spin electrons is partially linked to the dimerization in the atomic structure 
and in great part to the structure of deeper layers making the two
atoms in the chain inequivalent with respect to deeper layers, as shown in Fig. \ref{fig:magBands}.
The magnetic moments decrease of about one order of
magnitude between every two layers, going towards bulk diamond, as shown in Table \ref{table:spin}.

\begin{table*}[!htb] \footnotesize 
\caption{The magnitude of the spin of each atom in units of the Bohr
  magneton $\mu_B$ using different exchange and correlation functionals (XC)
  using the unrelaxed (dimerized) structure is taken from the non-magnetic calculations,
  and with the ion-relaxed and cell-relaxed calculations.
  The labeling of the atoms can be matched to Fig. \ref{fig:magBands}.
  }
\centering
\begin{ruledtabular}
        \begin{tabular}{l c c c c c c c c c} 
    XC & structure & 1 & 2 & 3 & 4 & 5 & 6 & 7 & 8 \\ 
        \hline
    PBE0  & unrelaxed    & -0.285(6) & +0.288(1) & +0.022(3) & -0.023(9) & -0.015(2) & +0.014(5) & +0.001(6) & -0.002(5) \\
    PBE0  & ion-relaxed  & -0.387(4) & +0.390(5) & +0.030(4) & -0.032(5) & -0.020(6) & +0.019(7) & +0.002(2) & -0.003(3) \\
    PBE0  & cell-relaxed & -0.374(8) & +0.377(9) & +0.031(4) & -0.033(5) & -0.021(8) & +0.020(8) & +0.002(1) & -0.002(9) \\
    HSE06 & unrelaxed    & -0.269(1) & +0.271(5) & +0.020(8) & -0.022(4) & -0.014(5) & +0.013(8) & +0.001(5) & -0.002(4) \\ 
    HSE06 & ion-relaxed  & -0.348(9) & +0.351(8) & +0.027(1) & -0.029(1) & -0.018(7) & +0.017(9) & +0.001(9) & -0.003(1) \\
	    \end{tabular}
 \end{ruledtabular}
\label{table:spin}
\end{table*}

As at the end of the magnetic simulation one of the atoms in the chain
experience a restoring force towards the non-dimerized
solution, we perform structural optimization in the presence of the
magnetic solution. 
We find that the spin-polarized structural
optimization restores the non-dimerized solution with a
negligible dimerization remaning on the surface atoms,
as shown in Table \ref{table:magstruc}.
Similarly, the dimerization of the atoms on the second layer, i.e. atoms 4 and 5,
also becomes negligible, while the bond lengths remain unchanged in the deeper layers.
We have also checked the effect of the cell relaxation with the PBE0 functional.
The in-plane lattice parameter decreases to $a=4.289$ \AA~ 
from the experimental value of $a=4.369$ \AA.
However, the conclusions after the ionic relaxation remain,
that the dimerization of the surface atoms is still negligible.

\begin{table}[!htb] \footnotesize
\caption{After inclusion of the magnetism, the bond lengths ($d_1$ and $d_2$) of the atoms in the Pandey $\pi$-chains in the
  topmost layer, dimerization $\Delta$, and buckling
  $\Delta z$ for different exchange and correlation functionals (XC).
  Unrelaxed refers to the dimerized structure obtained with the non-magnetic calculations.
  Changes in the structure is also presented after relaxing the ions only, as well as relaxing the whole cell for the PBE0 functional.}
\centering
\begin{ruledtabular}
        \begin{tabular}{l c c c c c}
        XC    & structure & $d_1$(\AA)& $d_2$(\AA) & $\Delta$ & $\Delta z$ (\AA) \\
        \hline
        PBE0  & unrelaxed    & 1.482(9) & 1.393(5) & 0.031(1) & 0.005(4) \\
        PBE0  & ion-relaxed  & 1.439(4) & 1.438(9) & 0.000(2) & 0.006(5) \\ 
	PBE0  & cell-relaxed & 1.419(8) & 1.419(7) & 0.000(1) & 0.006(7) \\
        HSE06 & unrelaxed    & 1.475(5) & 1.400(1) & 0.026(2) & 0.005(2) \\
        HSE06 & ion-relaxed  & 1.438(7) & 1.438(5) & 0.000(1) & 0.006(3) \\
        \end{tabular}
        \end{ruledtabular}
\label{table:magstruc}
\end{table}

Our calculations show that even in the absence of dimerization,
within hybrid functionals the ground state is a 
zero magnetization state, weakly ferrimagnetic on the two surface atoms,
as shown in Table \ref{table:spin}.
After the ionic relaxation, the magnetic moment of the atoms changes by $\sim 0.1~\mu_B$,
and the small imbalance on the magnetic moments of the surface atoms
remains to be $\sim 0.002-0.003~\mu_B$.
We have further checked the effect of the cell relaxation on the magnetic moments
using the PBE0 functional.
While the magnetic moment of the surface atoms decreased by $\sim 0.01~\mu_B$ with respect to the ion-relaxed calculation, 
the conclusion that the ferrimagnetism of the top surface atoms is stabilized remains.

\begin{table}[!htb] \footnotesize 
\caption{For the $\alpha$ (majority) and the $\beta$ (minority) bands,
			the fundamental band gap $E_g$, and the direct band gap at the high-symmetry points of the 
			Brillouin zone, J, K, $\Gamma$, J', given in eV,
			using the unrelaxed (dimerized) structure obtained from the non-magnetic calculations,
			as well as after ion and cell relaxation.					
                }
\centering 
\begin{ruledtabular}
        \begin{tabular}{l c c c c c c c }
        bands & structure    & spin  & $E_g$ & J     & K     & $\Gamma$ & J' \\
        \hline 
	PBE0  & unrelaxed    & $\alpha$ & 1.756 & 2.183 & 2.303 & 6.333 & 7.387 \\
	PBE0  & ion-relaxed  & $\alpha$ & 1.789 & 2.191 & 2.381 & 6.332 & 7.401 \\
	PBE0  & cell-relaxed & $\alpha$ & 1.711 & 2.174 & 2.363 & 6.154 & 7.424 \\
	HSE06 & unrelaxed    & $\alpha$ & 1.059 & 1.483 & 1.589 & 5.601 & 6.656 \\
	HSE06 & ion-relaxed  & $\alpha$ & 1.087 & 1.493 & 1.653 & 5.599 & 6.665 \\
	PBE0  & unrelaxed    & $\beta$  & 1.645 & 2.146 & 2.083 & 6.343 & 7.385 \\
	PBE0  & ion-relaxed  & $\beta$  & 1.638 & 2.135 & 2.088 & 6.345 & 7.398 \\
	PBE0  & cell-relaxed & $\beta$  & 1.508 & 2.060 & 2.022 & 6.163 & 7.462 \\
	HSE06 & unrelaxed    & $\beta$  & 0.943 & 1.299 & 1.369 & 5.611 & 6.655 \\
	HSE06 & ion-relaxed  & $\beta$  & 0.939 & 1.433 & 1.373 & 5.612 & 6.665 \\
        \end{tabular}
        \end{ruledtabular}
\label{table:magEg}
\end{table}

The weak ferrimagnetism of the two surface atoms is visible also in the electronic structure as
it splits the degeneracy of the spin bands in $\alpha$ (majority) and $\beta$ (minority) spin bands.
Table \ref{table:magEg} shows the band gap values at the high-symmetry points of the Brillouin zone
with the magnetic calculations.
The optical direct band gap between equal spin states is now 
$\approx 2.174 (2.060)$ eV for majority(minority) spins at the J-point with the PBE0 functional.
Finally, if we add the excitonic effects of $0.1-0.4$ eV \cite{Marsili2008} to the experimental gap of 1.47 eV, 
this leads to an experimental single particle gap of $1.57-1.87$ eV. 
The PBE0 gap value with the magnetism is thus on the higher side of the experimental window. 
With the HSE06 functional, the direct band gap at the J-point
is lower 1.493(1.433) eV for majority (minority) spin electrons,
hence on the lower side of the experimental window.

As the magnetic solution is a mean field solution, it is in principle
possible that the full many-body multi-determinant ground state is
actually non-magnetic, particularly in low dimensional materials. A many-body 
multi-determinant calculation with inclusion of structural optimization
is not possible for this system. Thus, in order to test the
reliability of hybrid functionals in predicting the competition
between dimerization and magnetism,  we consider carbyne,
the linear carbon chain that is a prototype of dimerization in one dimension.
This system is very pathological for what concerns structural and magnetic
instabilities and is known to be non-magnetic, but dimerized.
Our calculations show that the HSE06 functional favors the polyyne 
(dimerized carbyne) structure rather than cumulene (non-dimerized carbyne) 
structure -- unlike the standard LDA/PBE functionals.
This is in agreement with the literature \cite{carbyne}.
Furthermore, we have started with an initial AFM
configuration on the two atoms and found that the final state is non-magnetic.
Therefore, we show that in a similar carbon-based system, magnetism is not
stabilized even when we use a hybrid functional. Hence the magnetism in
diamond(111) surface can be a physical effect and our calculations predict that 
the magnetic solution is the most stable with a 
slight energy difference of a few meV/cell 
from the non-magnetic solution as shown in Table \ref{table:deltaE}.
Further experiments are needed to verify this hypothesis.

Our work demonstrates that, within hybrid functionals, the ground state of the diamond(111) 
surface is then insulating with zero net magnetization and
 ferrimagnetic order along the top surface atoms of the Pandey $\pi$-chains;
a very surprising result given that diamond is non-magnetic and the
atomic orbitals forming the surface state are of $p$ character and
thus, at the atomic level, not as localized as 3$d$ orbitals.

\section{Conclusion}

We have studied the diamond(111) surface by using hybrid-functionals with
different degrees of screened exchange. Contrary to all previous
theoretical works, we include the exchange interaction at all levels
in the calculation, both in the structural optimization
and in the calculation of electronic and properties.
Moreover, we allowed for magnetism in calculations.
 
In the absence of spin polarization, the primary effect responsible for the gap opening is the
dimerization of the Pandey $\pi$-chains, that is enough to lead
to an insulating state. This is at odds with all previous spinless calculations that were either
finding no gap \cite{Kresse1996,Bechstedt1996,Bechstedt2001,Stekolnikov2002}, or 
claimed that gap opening was purely an
electronic mechanism \cite{Marsili2005,Marsili2007,Marsili2008}. 
The PBE0 band gap of 1.672 eV is on the higher side, 
while the HSE06 band gap of 1.006 eV is on the lower side of the experimental
window for single particle gap of 1.57-1.87 eV obtained by summing
excitonic effects to the experimental gap. Thus, in the absence of spin
polarization the system could be classified as a Peierls-Slater insulator. 

By including spin polarization, we find that the flatness of the
diamond(111) 
edge-state stabilizes an insulating state with zero net magnetic moment and with ferrimagnetic ordering 
along the top surface atoms of the chain with sizable magnetic moments of the order of
$0.2-0.3~\mu_B$.
As the magnetic moment depends weakly on the 
underlying crystal structure, the electronic structure depends weakly
on the amount of dimerization. Interestingly, structural optimization
in the presence of magnetism converges to a ground state with
 a negligible dimerization on the surface atoms. 
We find that the PBE0 gap of $\approx 2.174 (2.060)$ eV is on the higher side,
while the HSE06 gap of 1.493(1.433) eV for majority (minority) spin electrons 
is on the lower side of the experimental window. Thus, within a
hybrid functional approach the ground state is essentially antiferromagnetic with negligible
dimerization, i.e. a Slater insulator.

As diamond(111) can be seen as formed from buckled graphene layers with
rhombohedral (ABC) stacking (see Fig.  \ref{fig:struc}), it is instructive to compare our
magnetic state with the one recently detected in multilayer graphene
with ABC stacking \cite{Lau2016,Thomas2015,Faugeras2016,Henck_Rhombo,BetulRhombo2017}.
In ABC graphene multilayers the state is globally antiferromagnetic,
but weakly ferrimagnetic on the outer layers, exactly as in the
present case. However, the magnetic moment per carbon atom is much smaller
than in diamond(111). It is, however, important to recall that the flat
edge states in ABC graphene extends in an extremely small part of the
Brillouin zone and hosts less electrons than the flat band of diamond(111). 
The similarity of these two states suggests the occurrence of a magnetic state even
in diamond(111). 

The fundamental point underlined by our work is that there are two
competing mechanisms for opening of a
gap in diamond(111), namely magnetization or dimerization.
Experimentally, it would be possible to detect the occurrence of 
magnetism via spin-resolved scanning tunneling spectroscopy by using
magnetic tips. 

Finally, our work demonstrates that in order to describe the correlated
states in flat edge bands, it is necessary to include the
electron-electron interaction at all levels in the calculations, both
in the structural and electronic properties. This is relevant far
beyond the case of diamond(111), and it is most likely also crucial to describe 
the phase diagram of twisted bilayer graphene\cite{CaoMott,CaoSC}, or
other low dimensional system presenting edge states such as IVB atoms
on top of IVB(111) surfaces such as
Pb/Si(111)\cite{PhysRevLett.120.196402} or one dimensional
polyenes\cite{Gu2017}.

\begin{acknowledgements}

This work is supported by the Graphene Flagship, by RhomboG grant (ANR-17-CE24-0030) 
from Agence Nationale de la Recherche and by PRACE.
Calculations were performed at IDRIS, CINES, CEA, BSC TGCC,
and CAC(Tardis).
B. P. acknowledges National Science Foundation 
(Platform for the Accelerated Realization, Analysis, and Discovery of Interface Materials (PARADIM)) 
under Cooperative Agreement No. DMR-1539918.
\end{acknowledgements}

\appendix
\section{Band gap convergence with respect to k-points}
\label{app:kpnt}

\begin{figure}[!thb]
        \centering
                \includegraphics[clip=true, trim=0mm 15mm 55mm 130mm, width=0.48\textwidth]{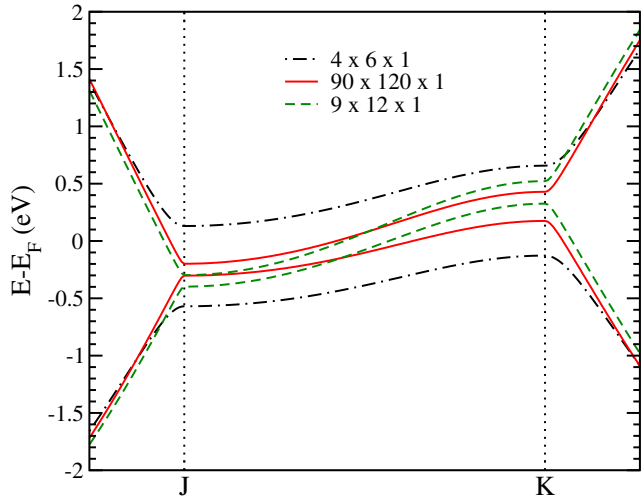}
        \caption{The electronic momentum (k-point) mesh convergence for the PBE0@PBE structure without magnetism.
        }
        \label{fig:kpnt}
\end{figure}

We test the convergence of the band gap with respect to the electronic momentum k-point mesh 
shown in Fig. \ref{fig:kpnt}.
For this purpose, we perform electronic structure calculations with the PBE0 functional on top of the
undimerized structure obtained with the PBE functional (PBE0@PBE).
The band gap can be similarly compared with the second panel of Fig. \ref{fig:allBands} (HSE06@PBE) 
of the manuscript.
Our tests show that the band gap strongly depends on the k-point grid.
A coarse k-point mesh of $4\times6\times1$ clearly overestimates the band gap 
with respect to the denser k-point meshes. This mesh is comparable
with that used in previous GW calculations\cite{Marsili2005,Marsili2007}.

\section{Comparison with ARPES data}
\label{app:arpes}

\begin{figure}[!thb]
        \centering
                \includegraphics[clip=true, trim=8mm 20mm 25mm 25mm, width=0.48\textwidth]{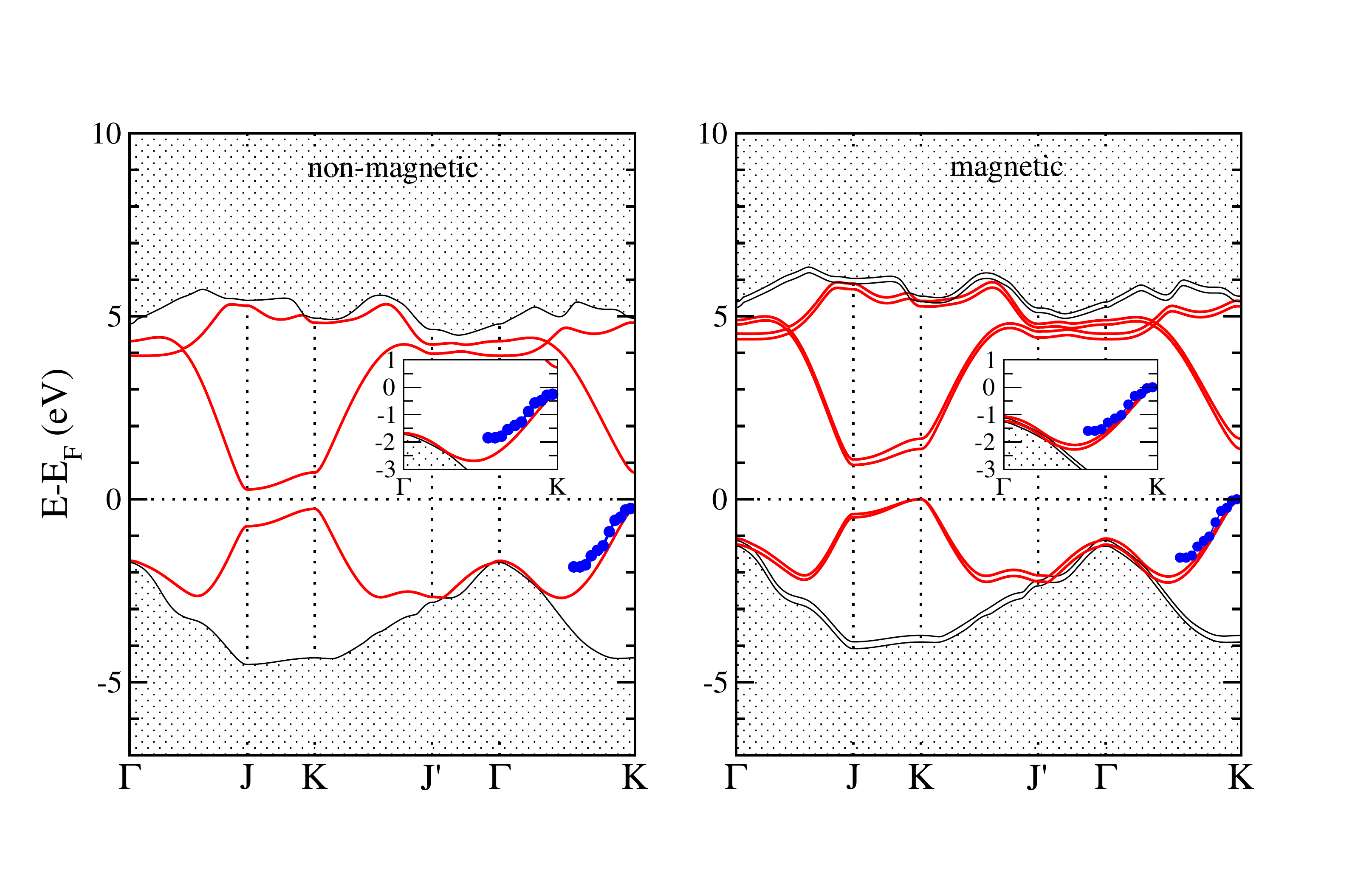}
        \caption{The electronic band structure of the surface states of diamond C(111)2$\times$1 surface,
        calculated with the HSE06 functional.
       Blue dots are the experimental ARPES data from Ref. \onlinecite{Stampfl1997ARPES}.
       The ARPES data are shifted such that the valence band maximum of calculation and experiment match at the K-point. 
        }
        \label{fig:hseBands}
\end{figure}

\begin{figure}[!thb]
        \centering
                \includegraphics[clip=true, trim=25mm 20mm 40mm 25mm, width=0.48\textwidth]{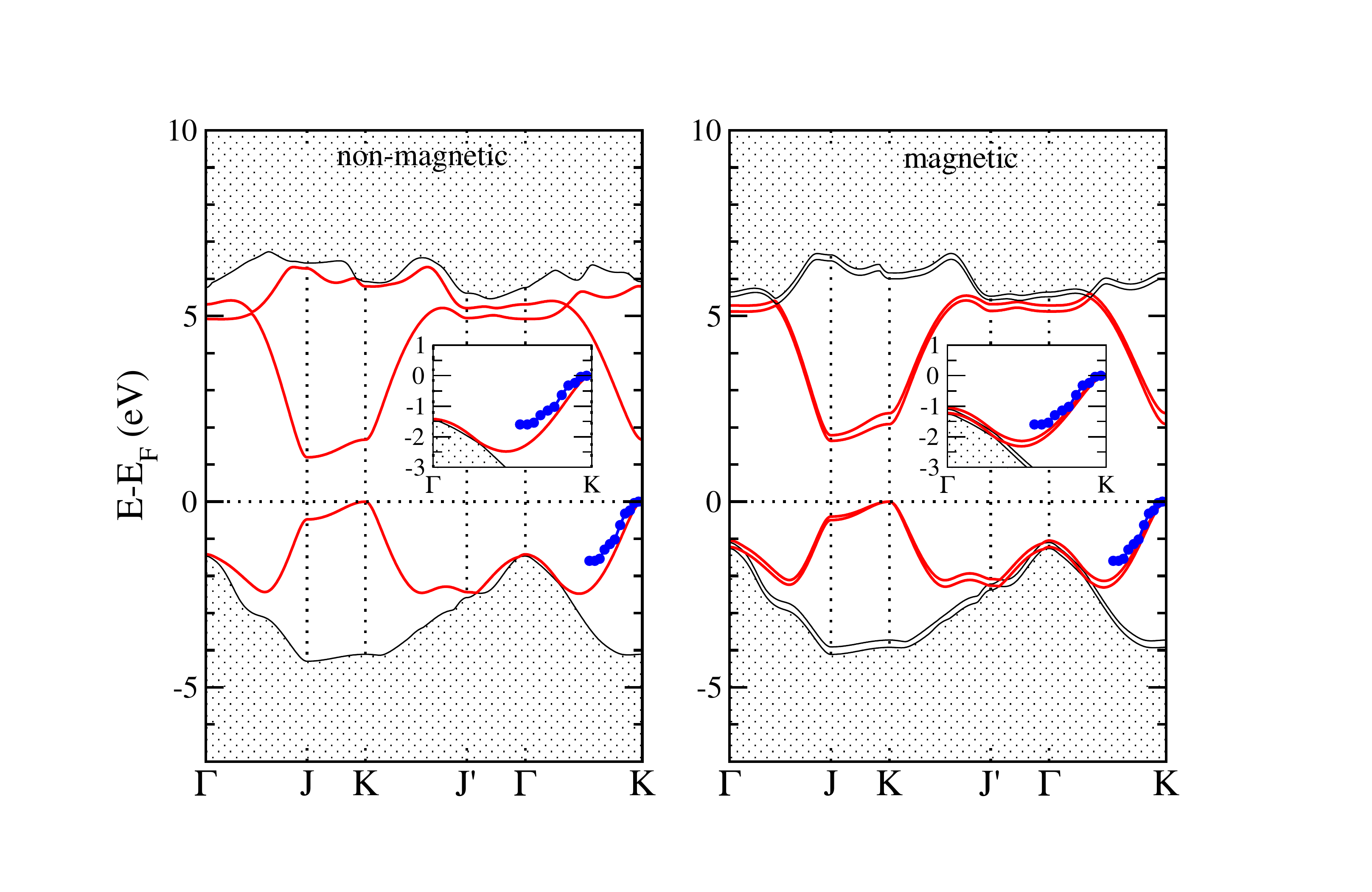}
        \caption{The electronic band structure of the surface states of diamond C(111)2$\times$1 surface,
        calculated with the PBE0 functional.
       Blue dots are the experimental ARPES data from Ref. \onlinecite{Stampfl1997ARPES}.
       The ARPES data are shifted such that the valence band maximum of calculation and experiment match at the K-point. 
        }
        \label{fig:pbe0Bands}
\end{figure}

We calculate the electronic band structure using the HSE06 and PBE0 functionals.
The electronic structure along the high-symmetry points of the Brillouin zone
is given in Fig. \ref{fig:hseBands} and Fig. \ref{fig:pbe0Bands}, respectively.
The valence band maximum is at the K-point and the conduction band minimum is at the J-point.
The experimental data are taken from the ARPES measurements \cite{Stampfl1997ARPES}.
We calculate the full path both for the non-magnetic and magnetic calculations with ionic relaxations.

The slope of the calculated path from the K-point towards the $\Gamma$-point 
depends on the amount of exchange and on the range of the interaction. 
For the PBE0 functional the best agreement is found with inclusion of magnetism.

\newpage

\bibliography{PaperBetul}

\end{document}